\documentclass[%
 reprint,
showpacs,preprintnumbers,showkeys,
 amsmath,amssymb,
 aps,
]{revtex4-1}

\usepackage{graphicx}
\usepackage{dcolumn}
\usepackage{bm}

\usepackage{natbib}

\usepackage{color}

\newcommand{\sref}[1]{Sec.~\ref{#1}}
\newcommand{\eref}[1]{Eq.~(\ref{#1})}
\newcommand{\esref}[1]{Eqs.~(\ref{#1})}
\newcommand{\fref}[1]{Fig.~\ref{#1}}

\begin{document}

\preprint{APS/123-QED}

\title{Solitary and Coupled Semiconductor Ring Lasers as Optical Spiking Neurons}

\author{W. Coomans}
\email[Electronic address: ]{wcoomans@vub.ac.be}
\author{L. Gelens}
\author{S. Beri}
\author{J. Danckaert}
\author{G. Van der Sande}
\affiliation{Applied Physics Research Group (APHY), Vrije Universiteit Brussel, Pleinlaan 2, B-1050 Brussel, Belgium}

\date{\today}

\begin{abstract}
We theoretically investigate the possibility of generating pulses in an excitable (asymmetric) semiconductor ring laser (SRL) using optical trigger pulses. We show that the phase difference between the injected field and the electric field inside the SRL determines the direction of the perturbation in phase space. Due to the folded shape of the excitability threshold, this has an important influence on the ability to cross it. A mechanism for exciting multiple consecutive pulses using a single trigger pulse (i.e. multi pulse excitability) is revealed. We furthermore investigate the possibility of using asymmetric SRLs in a coupled configuration, which is a first step toward an all-optical neural network using SRLs as building blocks.
\end{abstract}

\maketitle


\section{Introduction\label{sec:Introduction}}
Excitability is a phenomenon which is observed in a wide range of systems and has first been coined in biology to describe the behavior of individual nerve cells. Excitable behavior is, however, by no means limited to nerve cells. Other examples include nonlinear chemical reactions, cardiovascular tissues, ion channels, climate dynamics and lasers \cite{Lindner_PhysRep_2003}.
Common to all these excitable systems is their highly nonlinear response to external perturbations.
When unperturbed, the system remains quiescent and resides in a resting state. Small perturbations only lead to a small-amplitude linear response.
However, if the perturbation is sufficiently large, the system is transferred from the resting state to an excited state (the firing state). After this strong response, the system returns to its initial resting state through a refractory cycle. This large excursion of the system's variables in phase space corresponds to the emission of a large amplitude pulse. During the refractory cycle it is impossible to generate a second pulse---the system does not respond to any external perturbation.

In optical systems, excitability has attracted much interest in recent years \cite{Giudici_PRE_1997,Yacomotti_PRL_1999,Dubbeldam_PRE_1999,Giacomelli_PRL_2000,Larotonda_PRA_2002,Wunsche_PRL_2002,Wieczorek_PRL_2002,Piwonski_PRL_2005,Goulding_PRL_2007,Kelleher_OL_2009}. It provides a way to generate well-defined optical pulses and opens up possibilities for optical neural networks, all-optical pulse reshapers and delay lines. Lasers with saturable absorber \cite{Dubbeldam_PRE_1999,Larotonda_PRA_2002}, optically injected lasers \cite{Wieczorek_PRL_2002,Goulding_PRL_2007,Kelleher_OL_2009} and lasers with optical or opto-electronic feedback \cite{Giudici_PRE_1997,Yacomotti_PRL_1999,Giacomelli_PRL_2000,Wunsche_PRL_2002,Piwonski_PRL_2005} have all been proposed as optical excitable units.

In Ref.~\cite{Beri_PLA_2010} we proposed a mechanism for excitability in systems with a weakly broken $\mathbb{Z}_2$-symmetry close to a Takens-Bogdanov bifurcation \footnote{The type of excitability under consideration is generic for $\mathbb{Z}_2$-symmetric systems close to a Takens-Bogdanov bifurcation and with sign $s=-1$, when the normal form of the Takens-Bogdanov bifurcation in a two-dimensional $\mathbb{Z}_2$-symmetric system is written as $\dot x=y$ and $\dot y=ax+by+sx^3-x^2y$ \cite{Kuznetsov}.}. 
As optical prototypes of such systems we used semiconductor ring lasers (SRLs), whose active cavity has a circular geometry. As a result SRLs can generate light in two opposite directions referred to as the clockwise (CW) and the counterclockwise (CCW) mode.
The convenient device properties of SRLs allow this optical excitable unit to be highly integrable and scalable \cite{Hill_Nature_2004,Krauss_ElectLett_1990}, allowing for fully integrated optical neural networks and all-optical devices.

It was experimentally shown that short deterministic pulses can be excited by noise in asymmetric SRLs \cite{Beri_PLA_2010} and their origin was explained as a noise-activated escape across a barrier in an asymptotic 2D phase space \cite{Gelens_EPJD_2010}. Breaking the $\mathbb{Z}_2$-symmetry was realized by bringing a flat facet fiber very close to one of the output waveguides, increasing the reflective coupling for one of the counterpropagating modes. Moreover, that output waveguide was pumped to increase the power which is reflected [see \cite{Beri_PLA_2010} for more details]. Hence, an asymmetry in the magnitude of the linear coupling between the counterpropagating modes was obtained.

From an application point of view it would be desirable to excite pulses in a deterministic way by injection of an external optical trigger, which is theoretically investigated in this article.
We will also investigate the possibility of forming an optical neural network using SRLs as building blocks. Optical neural networks are attractive because of the high degree of parallelism that can be achieved and the large optical bandwidth that allows for very fast processing \cite{Hill_IEEE_2002,Kravtsov_OE_2011}. The typical spike duration in asymmetric SRLs is of the order of 10 ns, which is 5 orders of magnitude faster than the biological ms timescale. Kravtsov et al. \cite{Kravtsov_OE_2011} have recently demonstrated an optical spiking neuron operating at GHz speed, but it requires a substantial setup. We feel that the ability of integrating the SRL based neuron on chip offers a clear advantage. These advantages in speed and size yield good perspectives as an artificial neural network.
It will be shown in this article that coupled asymmetric SRLs are able to excite pulses in each other, mimicking neuron functionality as optical spiking neurons.

This article is organized as follows. The modeling of the device---both the rate equation model and the asymptotic phase plane---is discussed in \sref{sec:model}, in which we also briefly review the excitability mechanism. In \sref{sec:triggering} we analyze the SRL response to optically injected pulses, revealing a sensitive dependence on the phase of the optical trigger signal and a mechanism to excite consecutive pulses using a single trigger pulse. The possibility of coupling the SRLs toward the realization of an optical neural network will be examined in \sref{sec:neuralnetwork}. Finally, in \sref{sec:conclusion} we summarize the results.

\section{Model\label{sec:model}}
We use a general rate equation model as proposed in \cite{Sorel_OptLett_2002} which assumes that the SRL operates in a single transverse and single longitudinal mode and can sustain two counterpropagating directional modes. It consists of two complex mean-field equations for the counterpropagating modes $E_{1}$ (clockwise) and $E_{2}$ (counterclockwise) and a third equation for the carrier density $N$.
Following \cite{Sorel_OptLett_2002} with a straightforward modification to account for the optical injection, we can write the following rate equations for an optically injected asymmetric SRL; neglecting spatial variations within the laser and adiabatically eliminating the medium's polarization dynamics:
\begin{subequations}
\label{eq:RateEq}
\begin{align}
\dot E_{1} &= \kappa \left( {1 + \mathrm{i}\alpha } \right)\left[g_{1} N - 1\right]E_{1}
-k_1e^{\mathrm{i}\phi_k}E_{2} + F(t) \\
\dot E_{2} &= \kappa \left( {1 + \mathrm{i}\alpha } \right)\left[g_{2} N- 1\right]E_{2} - k_2e^{\mathrm{i}\phi_k}E_{1}\\
\label{eq:RateEqN}
\dot N &= \gamma \left[ \mu - N - g_{1}N\left|E_{1} \right|^2 - g_{2}N\left|E_{2} \right|^2\right]
\end{align}
\end{subequations}
Here the dot represents differentiation with respect to time $t$, $E_1$ and $E_2$ are the slowly varying complex envelopes of the counterpropagating waves, $N$ is the carrier population inversion, $g_{1}=1 - s\left| E_{1} \right|^2 - c\left| E_{2} \right|^2 $ and $g_{2}=1 - s\left| E_{2} \right|^2 - c\left| E_{1} \right|^2 $ are the differential gains of the modes ($s$ and $c$ respectively model the self- and cross-saturation effects), $\mu$ is the renormalized injection current ($\mu=0$ at transparency and $\mu=1$ at lasing threshold), $\kappa$ is the field decay rate, $\gamma$ is the carrier decay rate and $\alpha$ is the linewidth enhancement factor.
The term $F(t)=E_i(t) e^{\mathrm{i}(\Delta t +\varphi)}/\tau_{in}$ represents the optically injected trigger pulses in the clockwise mode, where $\tau_{in}$ is the cavity round-trip time, $E_i(t)^2$ the power envelope of the injected pulse, $\Delta$ the detuning between the frequency of the injected field and the cavity resonance frequency and $\varphi$ represents a constant phase difference.
A detuning $\Delta>0$ corresponds to an injected frequency which is higher than the cavity resonance frequency.
The linear coupling between the counterpropagating waves, referred to as backscattering, is caused by reflections inside the cavity at the interface with the coupling waveguide and at the cleaved end facets of the output waveguide. It is modeled by a backscattering amplitude $k_i$ with a phase shift $\phi_k$. The backscattering phase $\phi_k$ is chosen to be identical for both counterpropagating modes since it has been shown that an asymmetry in $\phi_k$ has no influence on the topology of the phase space structure \cite{Gelens_EPJD_2010}. The asymmetry in the backscattering amplitude $\Delta k=k_2-k_1$ is defined by the dimensionless parameter $\delta=\Delta k/2k$ where $k$ is the average backscattering amplitude. This asymmetry causes the SRL to be excitable \cite{Beri_PLA_2010, Gelens_EPJD_2010}. When $\delta>0$ in \eref{eq:RateEq}, residence in the $E_{2}$ mode is favored, allowing for excitable pulses of the $E_{1}$ mode. 

In a typical SRL, the photon lifetime $\kappa^{-1}$ and the carrier lifetime $\gamma^{-1}$ are respectively of the orders 10~ps and 5~ns, yielding two different time scales in the system. The other parameters are fixed to realistic values $\alpha=3.5$, $s=0.005$, $c=0.01$, the average backscattering value $k=0.44~\mathrm{ns}^{-1}$, $\phi_k=1.5$ and $\tau_{in}=0.6$~ps \cite{Sorel_OptLett_2002}, unless mentioned otherwise. The value of the bias current $\mu=1.65$ is chosen slightly above the value for which alternate oscillations disappear in a fold of cycles \cite{Gelens_EPJD_2010}.

Although we numerically investigate excitability in asymmetric SRLs with the rate equation model given in \eref{eq:RateEq}, the excitability mechanism can be interpreted more easily in a reduced two-dimensional phase space. It has been shown that on time scales slower than the relaxation oscillations the dynamics of the SRL essentially take place in a two-dimensional phase plane. The dynamical behavior in this phase plane is described by an asymptotically reduced model which has been introduced in \cite{VanderSande_JPhysB_2008}, characterized by the variables $\theta$ and $\psi$.
The resulting asymptotic description of the SRL is valid on time scales slower than the relaxation oscillations and it has been shown that it is able to predict many of the experimentally observed SRL characteristics \cite{Beri_PLA_2010,Gelens_EPJD_2010,Gelens_PRL_2009,Beri_PRL_2008}.
But including optical injection in the asymptotic model makes it rather cumbersome and its validity can be argued for short pulses. However, projecting our simulation results from \eref{eq:RateEq} on the asymptotic phase plane will prove to be useful.

The two phase space variables $\theta \in[-\pi/2,\pi/2]$ and $\psi\in[0,2\pi]$ are defined by
\begin{subequations}
\label{eq:RedEqDef}
\begin{align}
\theta &\equiv 2\arctan\left(\frac{|E_{2}|}{|E_{1}|}\right) - \frac{\pi}{2},\\
\psi &\equiv \phi_{2} - \phi_{1}.
\end{align}
\end{subequations}
where $\phi_i\equiv\arg(E_i)$. Hence $\theta$ is a measure for the relative power distribution among the counterpropagating modes ($\theta=\pi/2$ if $|E_{1}|=0$, $\theta=-\pi/2$ if $|E_{2}|=0$ and $\theta=0$ if $|E_{1}|=|E_{2}|$) and $\psi$ is the relative phase difference between the corresponding electric fields. The phase space topology of the asymmetric SRL in the $(\theta,\psi)$ phase plane is shown in \fref{fig:PulseEx}(b).

The grey and white regions indicate the basins of attraction of the two stable states in the SRL---the CW and CCW state---which are quasi-unidirectional. In this case, the CCW state is favored. They are separated by the stable manifold of a saddle point indicated by S. The branches of the unstable manifold respectively spiral toward the CW and the CCW state.
Assume that the unperturbed SRL resides in the CCW state. A perturbation forcing it to cross both branches of the stable manifold will make the SRL relax back to the CCW state by turning around the CW state, generating a large amplitude pulse. Hence the \emph{excitability threshold} is defined by the stable manifold of a saddle point which is folded throughout the phase space and separates the basins of attraction of the counterpropagating states.

We stress that the same results apply for every dynamical system with an underlying $\mathbb{Z}_2$-symmetry close to a Takens-Bogdanov point providing the same sequence of bifurcations~[13] (for more information on the bifurcation scenario specific to SRLs, we refer to \cite{Gelens_EPJD_2010}).

\section{Optical triggering of pulses\label{sec:triggering}}
\begin{figure}
\begin{center}
\includegraphics[width=\columnwidth]{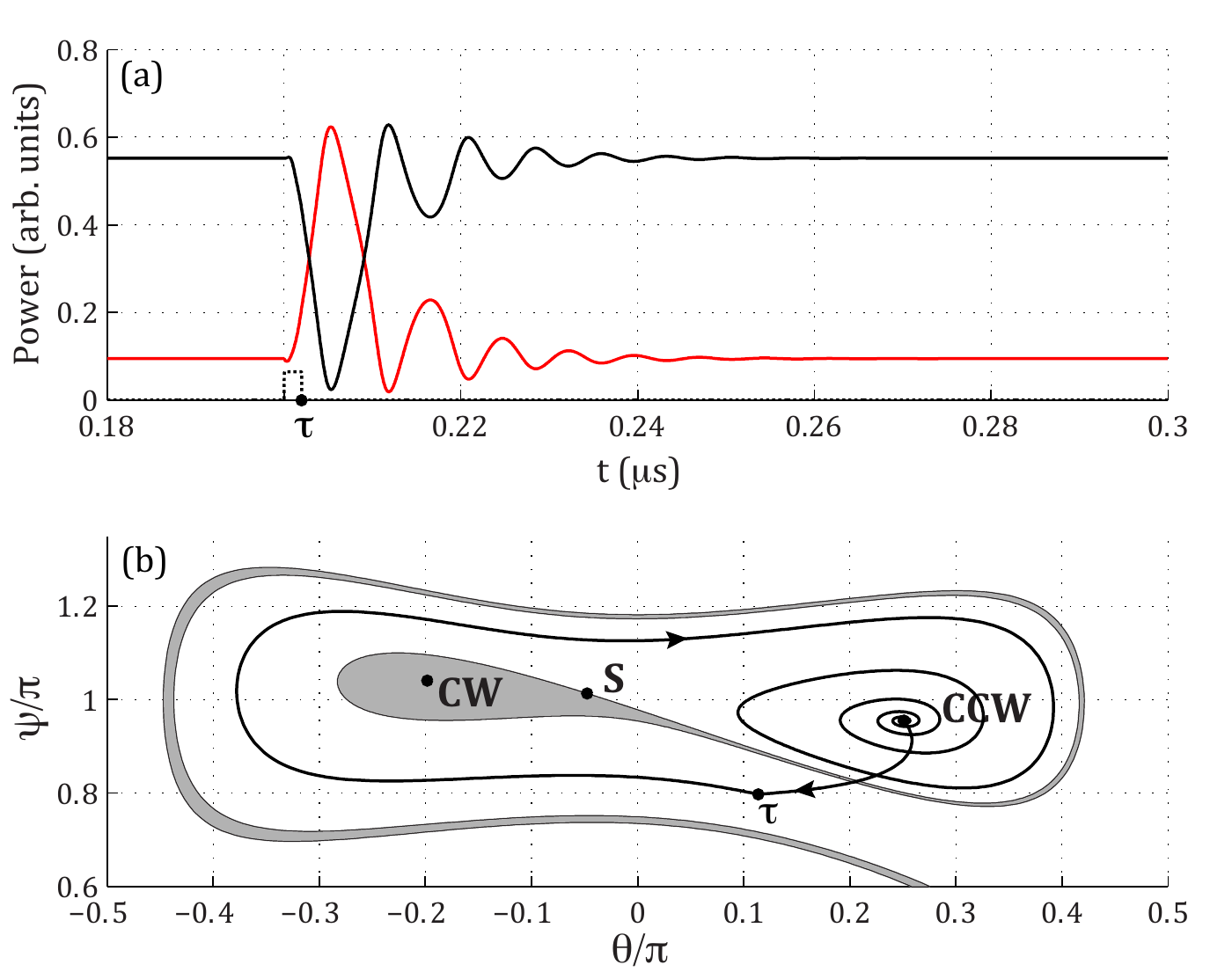}
\caption{(Color online) Simulation of \eref{eq:RateEq}  when optically injecting a 2 ns wide square pulse.
(a) Time trace of the modal intensities. The CW (CCW) modal power is depicted in red (black). The injected pulse is shown by a dashed black line. Note that the optically injected pulse power is scaled up by a factor 10$^7$ in this plot. $\tau$ indicates the time at which the injected pulse ends.
(b) Two-dimensional phase space trajectory corresponding to the time trace. The point $\tau$ also corresponds to the moment when the injected pulse ends. S indicates the location of the saddle. The basin of attraction of the CW (CCW) state is depicted in gray (white). Parameter values: $\delta=0.045$, $E_i=8\times10^{-5}$, resonant detuning, phase difference = $1.3\pi$.}
\label{fig:PulseEx}
\end{center}
\end{figure}
The asymmetric SRL will be excited to fire a pulse if a perturbation forces it to cross both branches of the stable saddle manifold by altering the phase difference $\psi$ and/or the relative power distribution $\theta$ between the counterpropagating modes. The stable manifold can be crossed by introducing spontaneous emission noise as shown in \cite{Beri_PLA_2010}. In this article we investigate the possibility of deterministically crossing the stable manifold by optically injecting a pulse into the SRL.
An example of SRL ``firing'' triggered by an optically injected pulse is shown in \fref{fig:PulseEx}(a). We have injected a 2 ns wide square pulse in the CW mode at resonant detuning (the optical frequency of the injected field and the SRL fields are identical). The SRL responds by emitting a large amplitude pulse in the CW mode, while the CCW mode acts in antiphase showing a power drop. Subsequently the SRL relaxes back to its former state in a decaying oscillatory fashion.

From the trajectory in the $(\theta,\psi)$ phase plane shown in \fref{fig:PulseEx}(b), we see that the trigger pulse kicks the SRL across both branches of the stable manifold of the saddle (grey region). When the trigger pulse is finished at the point $\tau$ the SRL relaxes back to the CCW state by turning around the CW state, which it is forced to do by the folded saddle manifold.
The oscillatory tail of the pulse corresponds to the spiraling movement towards the CCW state in the $(\theta,\psi)$ plane and indicates the existence of subthreshold oscillations. While decaying toward the resting state (CCW), the close proximity of the stable manifold allows for some perturbations to experience a temporary reduction of the distance to the excitability threshold, which has been shown to lead to noise-excited double pulses \cite{Beri_PLA_2010,Gelens_EPJD_2010}.

\subsection{Phase dependency}
It is clear from \fref{fig:PulseEx}(b) that the direction in which the SRL is kicked out of the CCW state has an influence on whether or not the stable manifold will be crossed, and hence whether or not the SRL will fire a pulse.

We examine the initial direction in phase space by looking at the optical injection terms in \eref{eq:RateEq}. Prior to the injection, the SRL will reside in the resting state for which the right hand sides of \eref{eq:RateEq} are zero in absence of the $F(t)$ term. Hence, at the start of the pulse the optical injection term is the only nonzero term.
For the square pulse injected in the CW mode, $F(t)=E_i e^{\mathrm{i}(\Delta t +\varphi)}/\tau_{in}$, and at the start of the pulse we get the following equations for the amplitude $|E_{1}|$ and the phase $\phi_{1}$ of the field of the CW mode:
\begin{subequations}
\label{eq:OptInj}
\begin{align}
\frac{d|E_{1}|}{dt}&=  \frac{E_i}{\tau_{in}} \cos (\phi_{1} - \Delta t_0-\varphi),\\
\frac{d\phi_{1}}{dt} &= -\frac{1}{|E_{1}|}\frac{E_i}{\tau_{in}} \sin (\phi_{1} - \Delta t_0-\varphi). 
\end{align}
\end{subequations}
Here $t_0$ indicates the time corresponding to the start of the injected pulse. Equations (\ref{eq:OptInj}) can easily be interpreted by considering either constructive ($\phi_{1} - \Delta t_0-\varphi=0$) or destructive ($\phi_{1} - \Delta t_0-\varphi=\pi$) interference.
\begin{figure}[tb!]
\begin{center}
\includegraphics[width=\columnwidth]{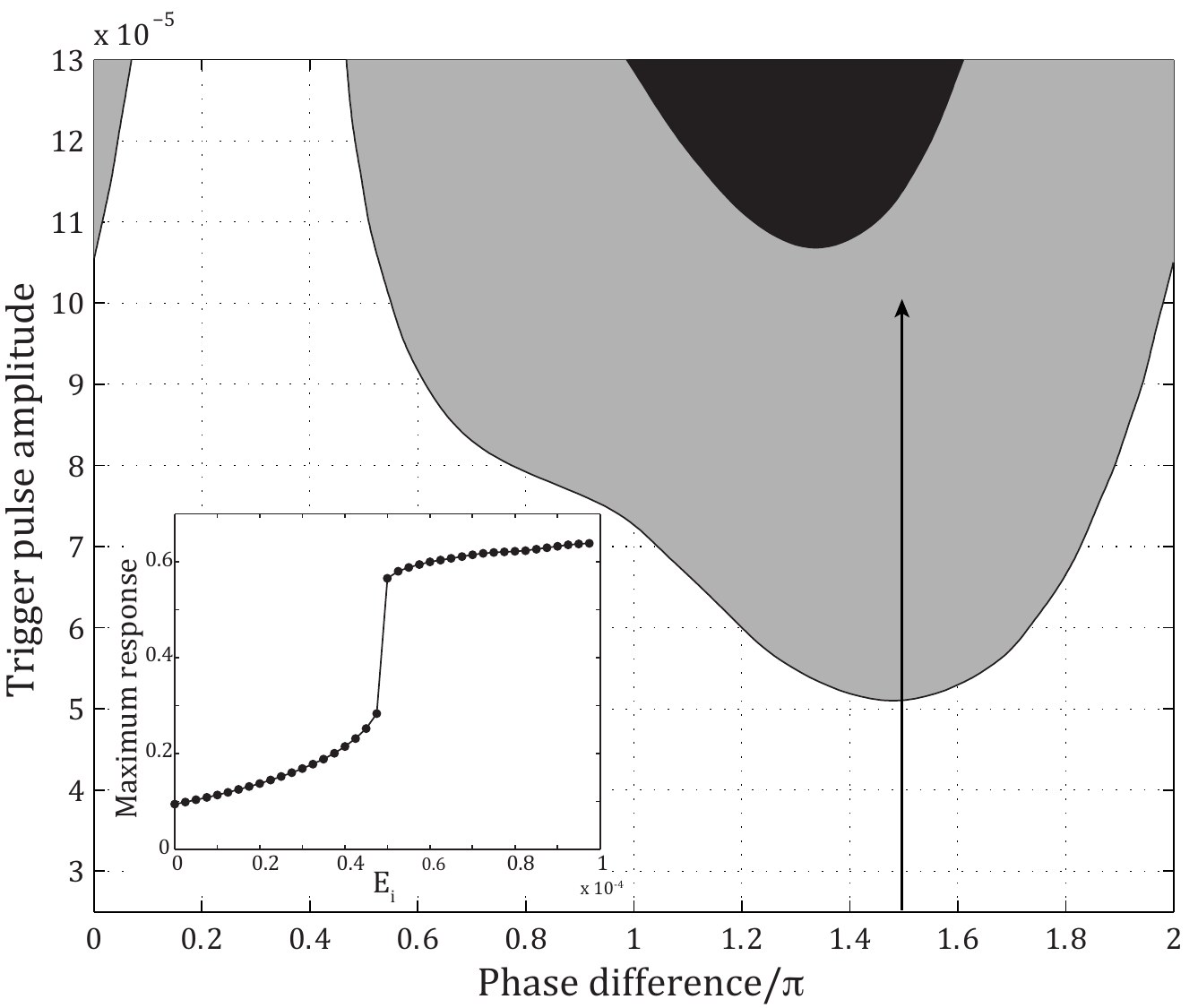}
\caption{(Color online) Influence of the phase difference between the injected field and the SRL field on the response of the SRL. A pulse with a fixed width of 2 ns is injected in the CW direction with varying phase difference and amplitude $E_i$.
Grey indicates excitation of a single pulse while black indicates excitation of two consecutive pulses. Inset: Magnitude of the response peak vs. amplitude of the optical excitation pulse for a fixed phase difference of $3\pi/2$ (corresponding to the arrow). Note the clear threshold behavior.}
\label{fig:PhaseDep}
\end{center}
\end{figure}
Reformulating \esref{eq:OptInj} for $\theta$ and $\psi$ using \esref{eq:RedEqDef} yields
\begin{subequations}
\label{eq:RedEqOptInj}
\begin{align}
\frac{d\theta}{dt} &= -\frac{2|E_{2}|}{|E_{1}|^2+|E_{2}|^2}\frac{E_i}{\tau_{in}}\cos (\phi_{1} - \Delta t_0-\varphi),\\
\frac{d\psi}{dt} &= \frac{1}{|E_{1}|}\frac{E_i}{\tau_{in}} \sin (\phi_{1} - \Delta t_0-\varphi). 
\end{align}
\end{subequations}
Equations (\ref{eq:RedEqOptInj}) show that the angle under which the trajectory leaves the CCW rest state is exactly the phase difference $\phi_{1} - \Delta t_0-\varphi$ (defined as the clockwise angle with the negative $\theta$ axis), which is an uncontrollable quantity in a practical setup.
This phase difference will play an important role in whether or not the SRL will be excited to fire a pulse. An unfavorable phase difference leads to a perturbation in the wrong direction and hence fails to excite the SRL. Based on the topology of the saddle manifolds in the asymptotic $(\theta,\psi)$ phase plane, we can predict which phase differences will have difficulties triggering a pulse.
Given that \esref{eq:RedEqDef} are only first order approximations with limited validity in time, and that the phase difference corresponds to the clockwise angle with the negative $\theta$ axis in the $(\theta,\psi)$ phase plane, a coarse estimate would be that phase differences between 0 and $\pi/4$ tend to be detrimental for triggering pulses.

This is confirmed in \fref{fig:PhaseDep}, showing the influence of the phase difference on the firing of the SRL by numerical simulation of the rate equation model in \esref{eq:RateEq}.
We inject a square optical pulse with a fixed width of 2 ns at resonant detuning and monitor the response.
In the diagram, both the amplitude $E_i$ and the phase difference between the injected field and the SRL fields are varied.
For low injection amplitudes, the SRL only fires a pulse when the phase difference is close to $-\pi/2$. In the $(\theta,\psi)$ plane this corresponds to a downward kick [comparable to \fref{fig:PulseEx}(b)], which is indeed the shortest way to cross the stable saddle manifold. For higher injection amplitudes, the phase condition is less stringent but nevertheless does not allow firing when the phase difference is close to $\pi/4$, which corresponds to our previous reasoning.
We can conclude that the phase difference between the fields determines the direction of the perturbation. Due to the folded shape of the excitability threshold in phase space (the stable manifold of the saddle point), it isn't accurate to speak of inhibitory and excitatory perturbations. But the folding is such that there does exist a ``wrong" direction for perturbations, along which no pulse will be excited [see \fref{fig:PhaseDep}].
The cumbersome phase dependency is an aspect that will arise in every optical excitable system, when triggered by an optically injected pulse.
The experimental results of W\"{u}nsche \emph{et al.} in \cite{Wunsche_PRL_2002}, covering an optically excitable multisection laser, indeed show a distinct jitter in the spikes although nominally identical pulses were injected, which is attributed to the phase difference between the fields.

 The inset in \fref{fig:PhaseDep} shows the distinct threshold behavior for raising injection amplitude when the phase difference is kept fixed. Below threshold, the response amplitude increases proportional to the trigger amplitude. Above threshold, it jumps to a much larger value and is less sensitive to the trigger amplitude. It was shown in \cite{Gelens_PRA_2010} that there exists an inverse correlation between the response pulse amplitude and the response pulse duration. The delay between the response and the trigger pulse ranges from  approximately 3 to 10~ns depending on the phase difference, since this determines where the threshold curve is crossed and hence also the length of the phase space trajectory up to  the pulse maximum.

\subsection{Multi pulse excitability}
\begin{figure}[t!]
\begin{center}
\includegraphics[width=\columnwidth]{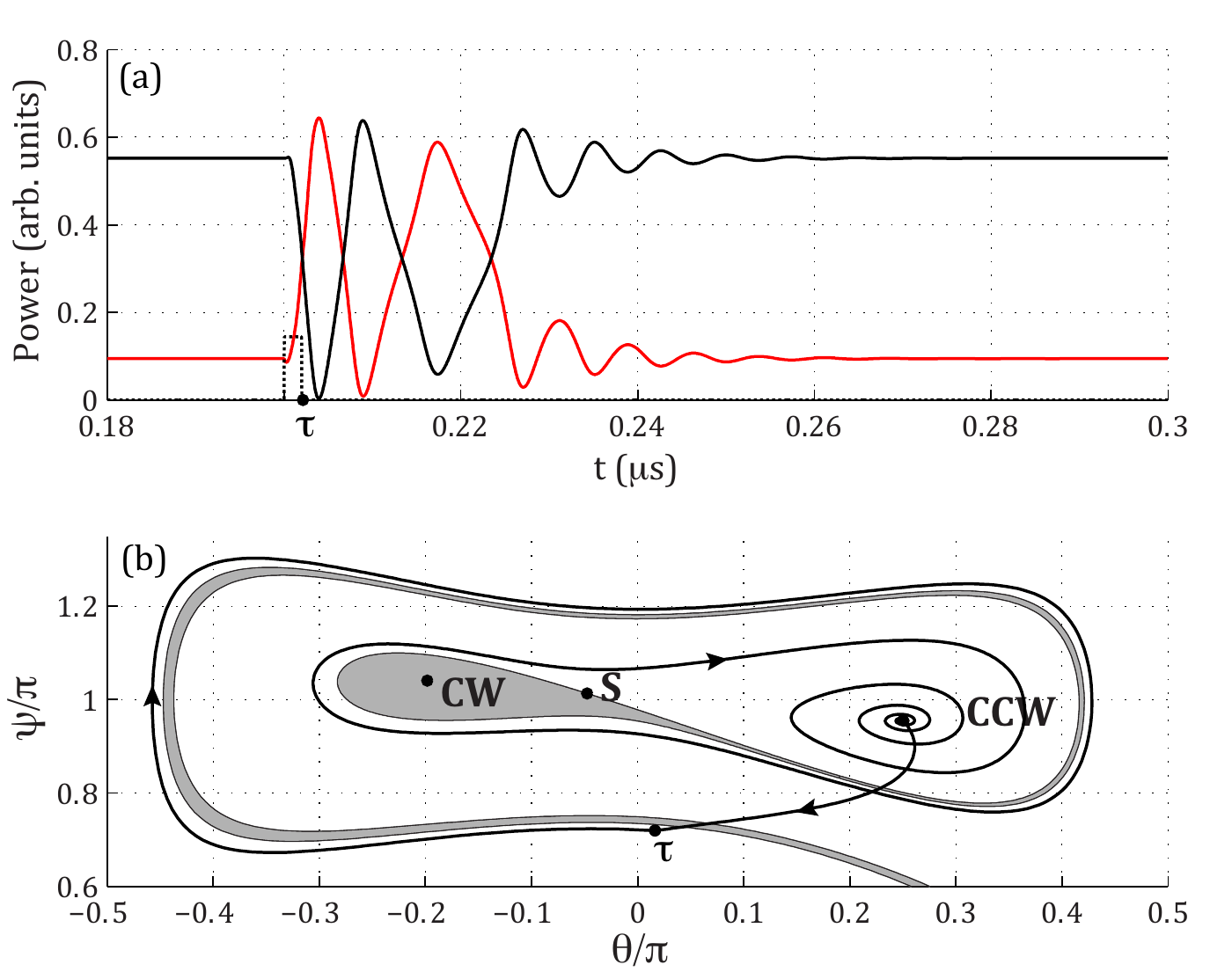}
\caption{(Color online) Simulation of \eref{eq:RateEq}  when optically injecting a 2 ns wide square pulse with a higher amplitude than in \fref{fig:PulseEx}.
(a) Time trace of the modal intensities.
(b) Two-dimensional phase space trajectory corresponding to the time trace. Parameter values: $\delta=0.045$, $E_i=1.2\times10^{-4}$, resonant detuning, phase difference = $1.3\pi$. Conventions as in \fref{fig:PulseEx}.}
\label{fig:DoublePulseEx}
\end{center}
\end{figure}

In the region of higher injection amplitude, the black area in \fref{fig:PhaseDep} indicates the excitation of two consecutive pulses, an example of which is shown in \fref{fig:DoublePulseEx}. Note that the threshold for this double pulse generation is formed by the second fold of the stable manifold. By changing the values of the pump current $J$ and the backscattering phase $\phi_k$ one can increase the number of folds of the stable manifold \cite{Beri_PLA_2010,Gelens_EPJD_2010}. This leads to identical scenarios in which more than two consecutive pulses can be excited using a single trigger pulse, i.e. multi pulse excitability. Multi pulse excitability was first reported for an optically injected laser by Wieczorek \emph{et al.} \cite{Wieczorek_PRL_2002}, explained by the vicinity of $n$-homoclinic bifurcations resulting in $n$ response pulses. Other reported mechanisms are the period-doubling of a limit cycle on which a saddle-node bifurcation takes place, resulting in double pulses \cite{Goulding_PRL_2007}, and slow-fast dynamics with folded slow manifold and a ramped parameter \cite{Wieczorek_PRSA_2011}.
The mechanism reported here is characteristic for systems with a phase portrait similar to \fref{fig:PulseEx}(b), i.e. $\mathbb{Z}_2$-symmetric systems close to a Takens-Bogdanov bifurcation~[13]. The multiple folding of the stable manifold, i.e. the threshold curve, around the resting state allows for a deterministic multi pulse return trajectory given that the perturbation is sufficiently large [see \fref{fig:DoublePulseEx}(b)].

Note that this multi pulse generation differs from the noise-induced clustering we previously reported in \cite{Beri_PLA_2010}. In that case the SRL is excited a second time while it is relaxing back to the CCW state close to the stable manifold, resulting in two very similar pulses. The deterministic return trajectory in \fref{fig:DoublePulseEx}(b) consists of a first, sharp, pulse followed by a wider pulse. The width of the excited pulses can vary depending on the strength of the initial perturbation, but the second pulse will always be wider than the first pulse. This range of widths is due to the characteristic flow in the reduced phase space \cite{Gelens_PRA_2010}.  The occurrence of this event in the noise-excited case \cite{Beri_PLA_2010} is very unlikely because of the the stochastic driving force and the invariant flow of the system.

\subsection{General considerations}
Since the excitability mechanism emerges from an asymptotic model which assumes the carrier density $N$ to be close to its threshold value, the scenario described in this article must assume the amplitude of the trigger pulse to be small [i.e. $O$($10^{-4}$)]. In this way, the SRL is confined on the $(\theta,\psi)$ subplane of its phase space [see Sec. \ref{sec:model}]. Injecting large pulses highly perturbs the value of $N$ and causes the SRL to leave this plane. Using such a large amplitude trigger pulse, it is hence possible to excite a pulse if the SRL happens to relax back to the the other side of the stable manifold on the $(\theta,\psi)$ plane. This excitability threshold is, however, obscured since the SRL dynamics can no longer be clearly described in the $(\theta,\psi)$ plane and the excited pulses have a distorted pulse shape due to large amplitude relaxation oscillations. The fact that the carrier density is not involved in the excitable excursion contrasts with the excitability mechanisms reported for other laser systems \cite{Dubbeldam_PRE_1999,Larotonda_PRA_2002,Wunsche_PRL_2002,Wieczorek_PRL_2002,Piwonski_PRL_2005,Goulding_PRL_2007}.
This is due to the ability of redistributing optical power between the counterpropagating modes, due to the backscattering $k$. In other laser systems one needs to generate additional optical power for the pulse for which the carrier reservoir needs to be addressed.

Besides the amplitude of the trigger pulse, the detuning between the optical frequency of the trigger pulse and the SRL fields also has an influence. 
In the case of resonant injection it is physically meaningful to talk about constructive or destructive interference between the injected field and the SRL field. However, if there is a frequency mismatch the interference will be constructive at one point, but destructive at another. This results in a periodic instead of constant driving force in the $(\theta,\psi)$ plane, which is unfavorable to cross the threshold (i.e. the stable manifold). Since the relevant physical quantity is the product $\Delta t$ of the frequency detuning $\Delta$ with time $t$, as long as $\Delta < O(1/\tau)$ (with $\tau$ the pulse width) the driving force will not vary significantly. In that case the scenario sketched above remains valid. This is also what we observe numerically.
We find numerically that there is a tolerance on the frequency mismatch of the order of a couple hundred MHz [cf. the pulse width is of the order of 2 ns], depending on the trigger pulse amplitude. There is of course a balance to be considered  between the trigger pulse amplitude and the detuning. When the frequencies of the injected field and the SRL field are largely detuned one can always use a shorter pulse with a higher amplitude, for which the previous considerations (of large pulse amplitudes) apply.

\section{Two coupled excitable SRLs\label{sec:neuralnetwork}}
\begin{figure}[t!]
\begin{center}
\includegraphics[width=\columnwidth]{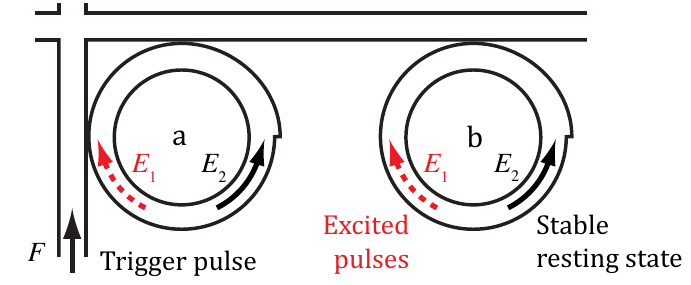}
\caption{(Color online) Schematic representation of the coupling scheme for excitable asymmetric SRLs considered in this article. The CW  (CCW) mode of SRL $a$ ($b$) is coupled in the CW (CCW) mode of SRL $b$ ($a$). The notch to the ring cavity is merely added as a visual indication of the asymmetry of the cavity. The curved black arrow indicates the stable resting state, the dashed red arrow indicates the propagation direction of excitable pulses. The injected trigger pulse ($F$) is shown by a straight black arrow.}
\label{fig:SetupCpld}
\end{center}
\end{figure}

In this section we will investigate whether excitable asymmetric SRLs are able to excite each other, paving the way to an integrated optical neural network. A detailed study about the dynamical behavior of coupled SRLs lies outside the scope of this work and will be dealt with elsewhere. We limit ourselves to excitable behavior and consider two SRLs coupled by a single bus waveguide as shown in \fref{fig:SetupCpld}.
In this type of coupling, the CW (CCW) mode of SRL $a$ ($b$) is coupled in the CW (CCW) mode of SRL $b$ ($a$). We therefore extend the model with two coupling terms:
\begin{subequations}
\label{eq:RateEqCpld}
\begin{align}
\dot E_{1a} &= \kappa \left( {1 + \mathrm{i}\alpha } \right)\left[g_{1a} N_a - 1\right]E_{1a}
-k_1e^{\mathrm{i}\phi_k}E_{2a} \notag\\&\qquad+F(t)\\
\dot E_{2a} &= \kappa \left( {1 + \mathrm{i}\alpha } \right)\left[g_{2a} N_a - 1\right]E_{2a} - k_2e^{\mathrm{i}\phi_k}E_{1a} \notag\\&\qquad- k_ce^{\mathrm{i}\phi_c}E_{2b}\\
\dot E_{1b} &= \kappa \left( {1 + \mathrm{i}\alpha } \right)\left[g_{1b} N_b - 1\right]E_{1b}
-k_1e^{\mathrm{i}\phi_k}E_{2b}\notag\\&\qquad - k_ce^{\mathrm{i}\phi_c}E_{1a} \\
\dot E_{2b} &= \kappa \left( {1 + \mathrm{i}\alpha } \right)\left[g_{2b} N_b - 1\right]E_{2b} - k_2e^{\mathrm{i}\phi_k}E_{1b}
\end{align}
\end{subequations}
The rate equation for the carrier density is given by Eq. (\ref{eq:RateEqN}) for both SRLs, with respectively $N=N_a$ and $N=N_b$. The term $F(t)$ again represents the external triggering and is only present in the equations for SRL $a$ [see \fref{fig:SetupCpld}]. Based on the results of the previous section we will use a trigger pulse with a favorable phase difference of $-\pi/2$ w.r.t. the fields in SRL $a$ [see \fref{fig:PhaseDep}].
The coupling is modeled by the coupling amplitude $k_c$ and the coupling phase $\phi_c$, which is the optical phase accumulated by the field when traveling from SRL $a$ to SRL $b$ (and vice versa). Note that Eqs. (\ref{eq:RateEqCpld}) are invariant under the transformation
\begin{subequations}
\label{eq:Transf}
\begin{align}
\phi_c &\mapsto \phi_c + \pi \\
\phi_{ia} &\mapsto \phi_{ia} \pm \pi/2 \\
\phi_{ib} &\mapsto \phi_{ib} \mp \pi/2 
\end{align}
\end{subequations}
where $\phi_{ij}\equiv\arg(E_{ij})$ and $i=\{1,2\}$. The second and third equation signify shifting the phase origin in each SRL over $\pi/2$, but in the opposite direction. The only difference with the untransformed system is that the value of the relative phase differences between the fields of the SRLs (without taking $\phi_c$ into account) will be shifted by $\pi$. However, the dynamical behavior of the global system will be identical. Therefore, we only need to consider $\phi_c\in[0,\pi]$ to grasp the behavior for all values of $\phi_c$.
All parameter values are identical to the previous section, except for the asymmetry $\delta$. We raise its value to 8\% in this section to increase the stability of the resting state. The phase space of the solitary SRL is qualitatively the same as sketched in the previous section up to $\delta=8.4$\%, where the saddle and the CW state disappear in a fold bifurcation [see Ref. \cite{Gelens_EPJD_2010} for more detailed information].
\begin{figure}[t!]
\begin{center}
\includegraphics[width=\columnwidth]{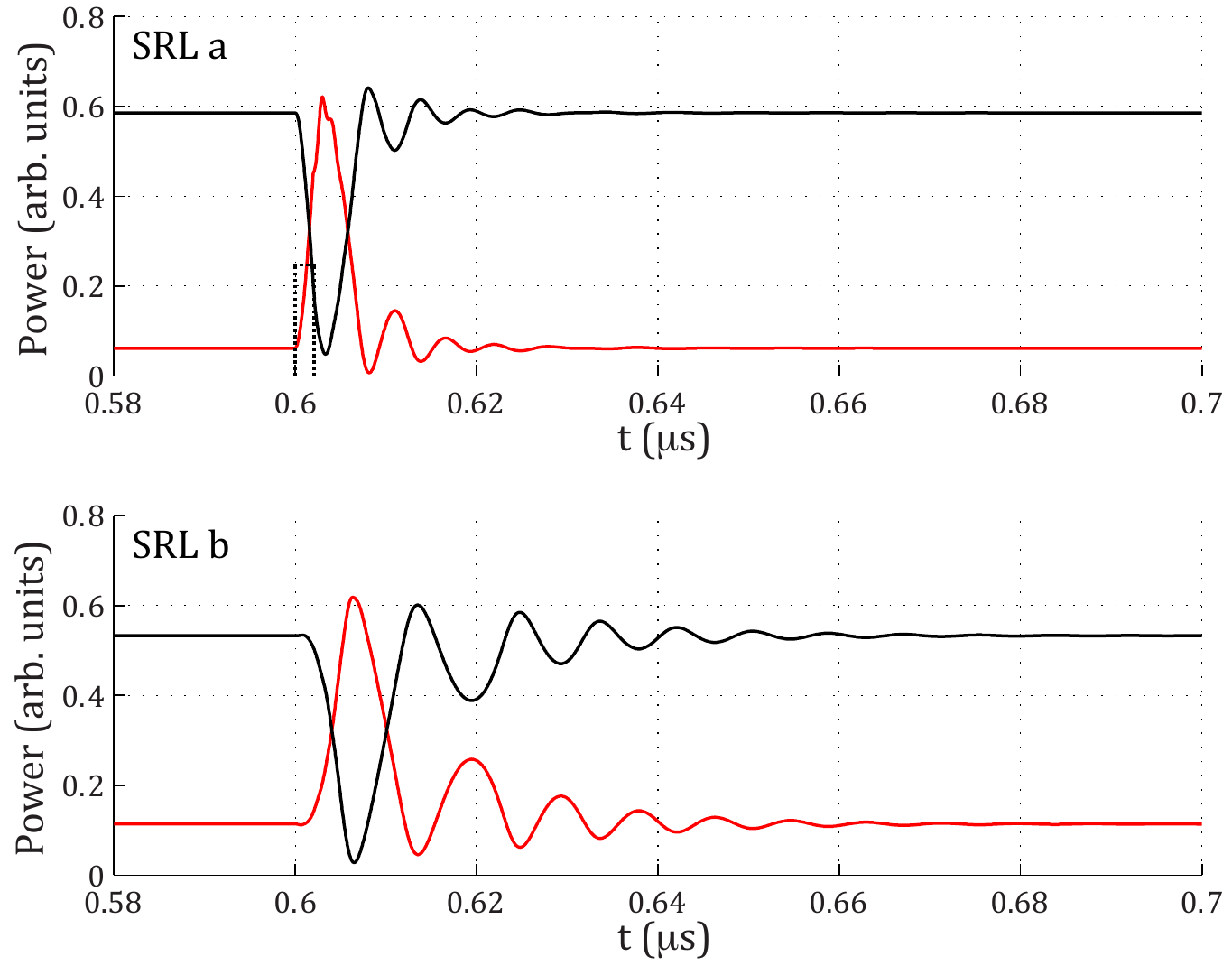}
\caption{(Color online) Simulated time traces of Eqs. (\ref{eq:RateEqCpld}) for the coupled setup shown in Fig. \ref{fig:SetupCpld}. The upper (lower) trace shows the powers in SRL $a$ ($b$). At $t=0.6~\mu\mathrm{s}$ a trigger pulse is injected solely in SRL $a$ [see Fig. \ref{fig:SetupCpld}]. Parameter values: $\delta=0.08$, $E_i=1.6\times10^{-4}$, $k_c=0.35k$, $\phi_c=0.35\pi$. Conventions as in \fref{fig:PulseEx}.}
\label{fig:CpldTT}
\end{center}
\end{figure}

For the coupling amplitude $k_c$ we have chosen $k_c<k$, which means that the reflective coupling inside each SRL has to be larger than the coupling to the other SRL. This is hard to achieve when the backscattering is small. However, the value of the backscattering increases with the reflective coupling between the counterpropagating modes, which can be manipulated through the cavity design. 
Due to the weak character of the coupling, excitatory excursions still persist due to the similar phase space structure. We will, however, not project the trajectories on the respective $(\theta,\psi)$ planes of each SRL. The phase space structure of the solitary SRL in these planes does not provide a suitable reference since the projection does not capture the extra dimension of the coupled system.

Coupling the two SRLs nevertheless has an effect on the dynamics of the individual lasers. Due to the single waveguide coupling SRL $a$ and $b$ are continuously injected in respectively the CCW and the CW mode, which stabilizes these respective modes in the SRLs \cite{Coomans_PRA_2010}. On the other hand, the cavity asymmetry $\delta$ favors the CCW mode in both SRLs. In the range of parameters we consider, the coupled system has a steady state $s_1$ that has both SRLs lasing in the CCW mode  [see \fref{fig:SetupCpld}],  which is stable throughout the considered parameter range. We raised the value of $\delta$ in this section to increase the stability of this initial steady state.
Furthermore, there is another stable steady state $s_2$ in which SRL $a$ and $b$ are respectively lasing in the CCW and CW mode. This steady state does not exist for coupling phases $\phi_c$ close to 0 (and $\pi$, due to the transformation in \esref{eq:Transf}).

\fref{fig:CpldTT} shows that a pulse generated in SRL $a$---excited by an external optical trigger signal---can in its turn excite a pulse in SRL $b$. Note that the pulse shape in SRL $a$ is a little distorted. Small oscillations are superimposed on it and the damping to the resting state is increased. The superimposed oscillations are relaxation oscillations as explained in the previous section for high(er) amplitude trigger pulses. The need for a slightly higher amplitude stems from the stabilizing effect of the continuous injection from SRL $b$, which also explains the increased damping rate toward the resting state. This change in damping rate is not very visible in SRL $b$ [compare with \fref{fig:PulseEx}] because the weakest mode of SRL $a$ is injected in the CW (firing) mode of SRL $b$. The different power levels of the resting state in SRL $a$ and $b$ in \fref{fig:CpldTT} is an artifact of the single waveguide coupling, which produces an additional asymmetry of the global system. The dominant mode of SRL $b$ is coupled to SRL $a$, while the weakest mode of SRL $a$ is coupled to SRL $b$. The dominant mode of SRL $a$ and the weak mode of SRL $b$ are not fed through. This gives rise to the asymmetric steady state power levels.
If we would add an additional bus waveguide connecting the bottom of both rings, the dominant mode of SRL $a$ and the weak mode of SRL $b$ would also be fed through. No additional asymmetry is introduced in this case and the steady state power levels are symmetric. This double waveguide coupling also yields positive results with regard to pulse excitation in SRL $b$, but exhibits much more unstable behavior than the single waveguide coupling. For this reason, we will focus on the single waveguide coupling scheme of \fref{fig:SetupCpld}.

\begin{figure}[t!]
\begin{center}
\includegraphics[width=\columnwidth]{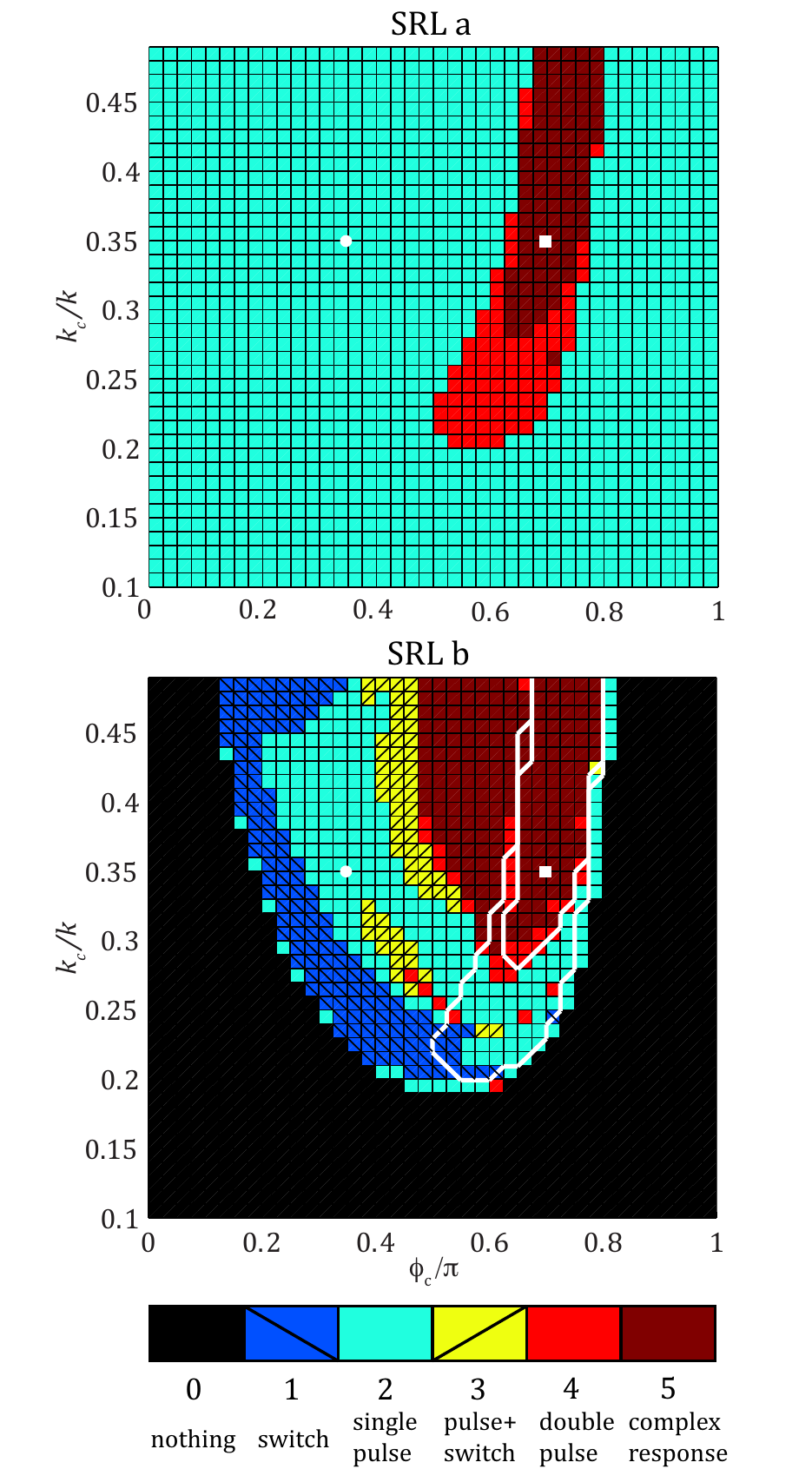}
\caption{(Color online) The response of SRL $a$ (top) and $b$ (bottom) for different values of $k_c$ and $\phi_c$. The colormap displays the number of times the line $\theta=0$ is crossed in the asymptotic $(\theta,\psi)$ phase plane. Number of crossings larger than 5 have been renormalized to 5 to improve the readability. The upper (lower) white line in the bottom figure bounds the region where the number of crossings in SRL $a$ is $\geq5$ ($=4$). The white circle (square) indicates the parameter values for the time trace in \fref{fig:CpldTT} (\fref{fig:CpldTTPhase}).
Trigger pulse: $E_i=1.6\times 10^{-4}$. Parameter values: $\delta=0.08$.}
\label{fig:CpldScan}
\end{center}
\end{figure}

The response of SRL $a$ and $b$ for various values of $k_c$ and $\phi_c$ are shown in \fref{fig:CpldScan}. The color coding represents the number of times the line $\theta=0$ is crossed in the asymptotic $(\theta,\psi)$ phase plane, which corresponds to an equal power in both modes. Two subsequent crossings in SRL $b$ can hence be regarded as an excited pulse. An odd number of crossings signifies that SRL $b$ has switched from CCW to CW lasing, instead of returning to its resting state and emitting a pulse.
For the coupled system this indicates a switch to the steady state $s_2$. This is the case for the pale blue (1 crossing), yellow (3 crossings) and part of the brown area (5 or more crossings) in \fref{fig:CpldScan}.

\begin{figure}[t!]
\begin{center}
\includegraphics[width=\columnwidth]{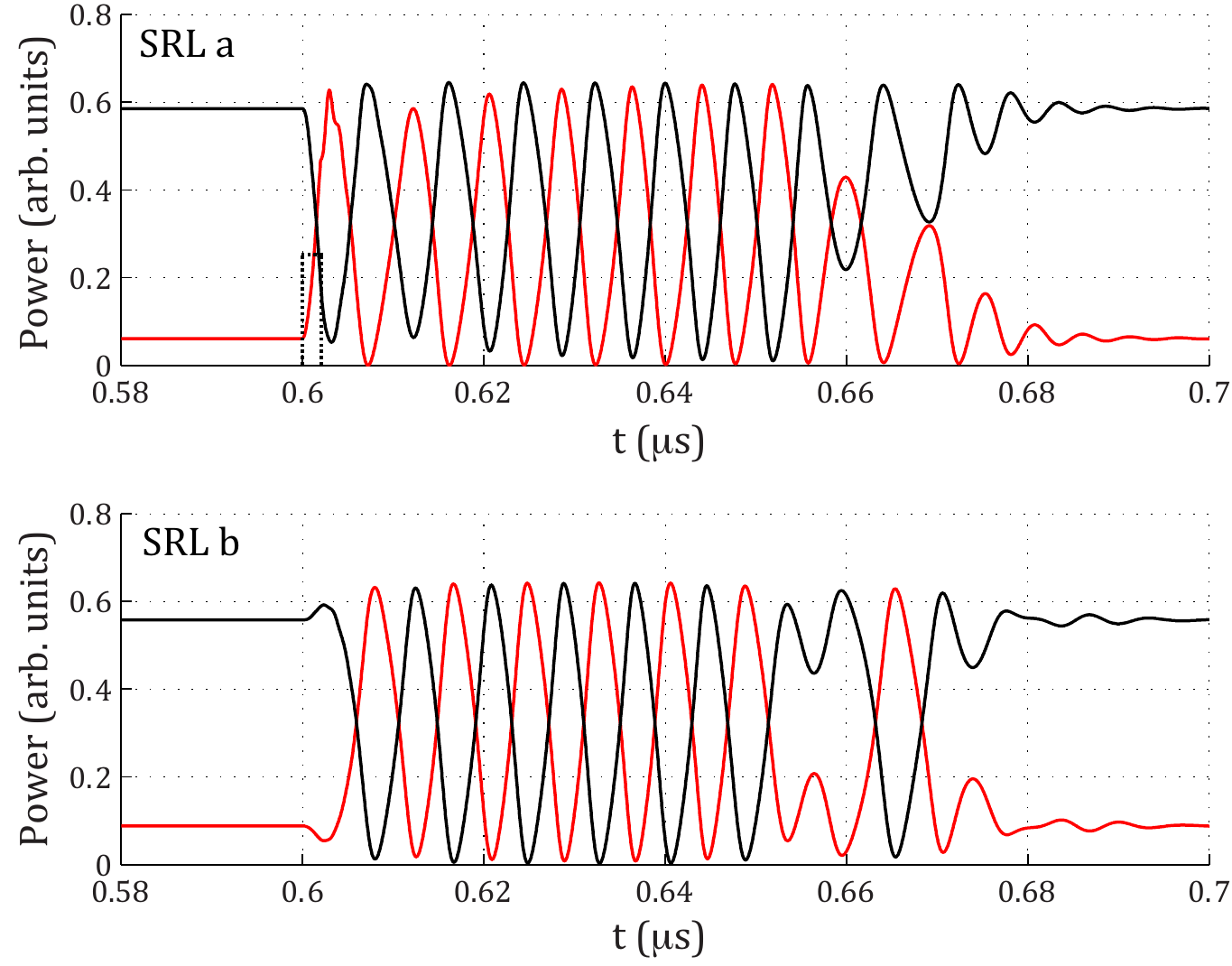}
\caption{(Color online) Simulated time traces of Eqs. (\ref{eq:RateEqCpld}) for the coupled setup shown in Fig. \ref{fig:SetupCpld}. The upper (lower) trace shows the powers in SRL $a$ ($b$). At $t=0.6\mu\mathrm{s}$ a trigger pulse is injected solely in SRL $a$ [see Fig. \ref{fig:SetupCpld}]. Parameter values: $E_i=1.6\times10^{-4}$, $\delta=0.08$, $k_c=0.35k$, $\phi_c=0.7\pi$. Conventions as in \fref{fig:PulseEx}.}
\label{fig:CpldTTPhase}
\end{center}
\end{figure}

Just as in the previous section, the optical phase plays an important role. Certain values of $\phi_c$ will be favorable and others will be detrimental for transferring excited pulses. The role of $\phi_c$ is however not the same as the role of the ``phase difference'' between the injected and the SRL fields in the previous section. When residing in their resting state SRL $a$ and $b$  will have a constant phase difference between their fields, which is a function of $\phi_c$. Moreover, the phase of the ``source'' laser (SRL $a$) is no longer independent of the injected laser due to the bidirectional character of the coupling.
Nevertheless, there is a region around $\phi_c=0$ (and $\phi_c=\pi$) for which no response can be elicited. Values of $\phi_c$ around $\pi/2$ yield clear responses from SRL $b$, i.e. the number of crossings is nonzero. The area with value 2, in which we excite a single pulse in SRL $b$, is relatively large. An example of a time trace in this area was shown in \fref{fig:CpldTT} and is indicated by a white circle in \fref{fig:CpldScan}. The brown region at $\phi_c\approx0.6\pi$ indicates more complex behavior than single pulse excitations, an example of which is shown in \fref{fig:CpldTTPhase}. The corresponding location in \fref{fig:CpldScan} is indicated by a white square. Note that both time traces have the same coupling amplitude but a different coupling phase. Instead of each firing a single pulse, the different value of $\phi_c$ allows SRL $a$ and $b$ to emit a series of spikes. Note that they are emitted in anti-phase by SRL $a$ and $b$. After a while this oscillation can no longer be sustained and the SRLs return to their resting states. 
The range over which both SRL $a$ and $b$ yield a complex response coincides with the existence of an unstable limit cycle. For increasing $\phi_c$ it originates in a homoclinic bifurcation on the left side of the complex region (the side of lower $\phi_c$), and is destroyed in a subcritical Hopf bifurcation on the right side of the complex region, resulting in the destabilization of $s_2$.
So different values of the coupling phase can have a profound influence on the dynamical behavior.

Two coupled asymmetric SRLs can thus function as communicating neurons. We moreover observe that they do this in a single direction, i.e. the configuration of Fig. \ref{fig:SetupCpld} enforces a preferential ``propagation'' direction for the excited pulses. We have shown that it is possible to excite a pulse in SRL $b$ with a 
pulse from SRL $a$ that was excited by an external trigger. However, when exciting a pulse in SRL $b$ with an external trigger, we have not found any parameter values for which SRL $b$ excites a pulse in SRL $a$.
Note that in this case the ``pulse'' traveling from SRL $b$ to SRL $a$ is actually not a pulse, but rather a dip in the steady state power. The pulse can actually be interpreted as an increase in a \emph{destabilizing} injection, while the dip is a decrease in a \emph{stabilizing} injection [see \fref{fig:SetupCpld}], explaining the difference in response sensitivity.

\section{Concluding remarks\label{sec:conclusion}}
In this work, we have theoretically investigated the possibility of triggering a pulse in an excitable (asymmetric) SRL by using an optically injected trigger pulse. We have used a standard rate equation model for the numerical simulations and an asymptotic two-dimensional phase plane to interpret the results. This two-dimensional phase plane provides a well-defined threshold appearing as two branches of a stable saddle manifold that need to be crossed.
Using this approach we have shown that the phase difference between the injected field and the SRL fields determines the direction of the perturbation in this phase plane. An unfavorable phase difference, which we were able to derive from the asymptotic phase plane structure, fails to excite a pulse in the SRL.
Furthermore, a mechanism for exciting two or more consecutive pulses using a single trigger pulse was revealed. The appearance of such multiple response pulses is not related to period-doubling or homoclinic bifurcations as in \cite{Wieczorek_PRL_2002,Goulding_PRL_2007}, nor to noise-induced clustering as in \cite{Beri_PLA_2010}. They arise due to the multiple folding of the stable manifold, i.e. the threshold curve, around the resting state.

We have also shown that two SRLs, coupled by a single bus waveguide, can excite each other and can thus function as communicating neurons. This type of neural network can be fully integrated on chip and does not suffer from the drawback of needing extra-cavity measures as other optical neurons do \cite{Giudici_PRE_1997,Yacomotti_PRL_1999,Giacomelli_PRL_2000,Wunsche_PRL_2002,Piwonski_PRL_2005,Dubbeldam_PRE_1999,Larotonda_PRA_2002,Wieczorek_PRL_2002,Goulding_PRL_2007,Kelleher_OL_2009}. The typical pulse duration of 10~ns moreover yields a processing speed that is $10^5$ times larger than biological neurons. These advantages in size and speed, and the high degree of parallelism offer good perspectives as an artificial neural network.

Although the asymptotic two-dimensional model used here is derived particularly for SRLs, the excitability mechanism presented in this work is general for a subclass of $\mathbb{Z}_2$-symmetric systems close to a Takens-Bogdanov point providing the same sequence of bifurcations~[13]. As an example of another optical system that will exhibit similar dynamics, we mention semiconductor disk lasers \cite{Liu_NatPhot_2010} as they essentially share the same circular symmetry as SRLs.

\begin{acknowledgments}
This work has been funded by the Research Foundation Flanders (FWO). This work was supported by the Belgian Science Policy Office under grant IAP-VI10 and the European project PHOCUS (EU FET-Open grant: 240763). W.C. is a PhD Fellow and  L.G., S.B. and G.V. are Postdoctoral Fellows of the FWO.
\end{acknowledgments}

%

\end{document}